\documentstyle[manuscript,aps]{revtex}
\begin{document}
\draft


\widetext

\title{Quasiparticle conductivities in disordered $d$-wave 
superconductors}
\author{Michele Fabrizio$^{a,b}$, Luca Dell'Anna$^a$, and Claudio 
Castellani$^c$}
\address{$^a$ Istituto Nazionale di Fisica della Materia and 
International School for Advanced Studies (SISSA)\\
Via Beirut 2-4, I-34014 Trieste, Italy}
\address{$^b$ International Center for Theoretical Physics (ICTP)\\
Strada Costiera 11, I-34100 Trieste, Italy}
\address{$^c$ Istituto Nazionale di Fisica della Materia and 
Universit\`a degli Studi di Roma ``La Sapienza'', 
Piazzale Aldo Moro 2, I-00185 Roma, Italy}
\date{\today}
\maketitle 
\begin{abstract}
We study the quasiparticle transport 
coefficients in disordered $d$-wave superconductors. We find that 
spin and charge excitations are generally localized unless magnetic 
impurities are present. If the system is close to a 
nesting point in the impurity-scattering unitary limit, 
the tendency towards localization is reduced while 
the quasiparticle density of states gets enhanced 
by disorder. We also show that the residual repulsive interaction 
among quasiparticles has a delocalizing effect and increases 
the density of states. 
\end{abstract}
\pacs{72.15.Rn,74.20.-z,74.72.-h,74.25.Fy}

\narrowtext

The discovery of $d$-wave superconductivity in doped cuprate materials 
has stimulated a substantial experimental and theoretical activity 
focused on the role played by disorder in superconductors with gapless 
quasiparticle excitations.

Indeed this type of superconductors raises several questions 
which are worth being addressed. For instance, charge is not 
anymore a conserved quantity in the superconducting phase, contrary to   
spin and energy. Therefore, while quasiparticle spin and thermal density 
long wavelength fluctuations are diffusive in the presence of disorder, 
charge density fluctuations do not diffuse, realizing a 
sort of {\sl spin-charge} separation in the hydrodynamic limit. 
As it has been shown in Ref.\cite{Fisher}, the effective field theory,   
which describes   
the spin and thermal diffusive modes, has still the form of a 
non linear $\sigma$-model, like in the standard Anderson localization. 
The $\beta$-function of the appropriate $\sigma$-model 
implies that in two dimensions the spin diffusion constant vanishes.
However, in view of the previously discussed independence between charge and spin 
long-wavelength excitations, spin and charge might behave differently
with respect to localization, resulting into  a 
quite intriguing scenario.

In this work we rexamine the role of quantum interference 
corrections to the transport properties of  
quasiparticles in a $d$-wave superconductor, including besides spin 
also quasiparticle charge conductivity. Moreover, to keep our analysis 
as general as possible,  
we consider several possible symmetries of the 
underlying model. Namely, we study models with and without time 
reversal 
invariance, as well as in the presence or absence of the so-called 
sublattice symmetry\cite{Gade}, which  
might be relevant close to a nesting point in the unitary 
scattering limit. 
In all the above symmetry classes, we derive the renormalization group 
(RG) equations 
for the charge and spin conductivities as well as for the density of states 
(DOS). We find that the spin and charge conductivities behave similarly 
in the absence of sublattice symmetry, and differently otherwise, in which 
case the spin conductivity stays finite while the charge one vanishes. 
The impurity induced DOS in the absence of sublattice symmetry is suppressed 
for potential scattering, in 
agreement with Ref. \cite{FisherDOS}, but enhanced in the presence of 
spin-flip scattering, which also gives a delocalizing correction to 
conductivities. On the contrary, if sublattice symmetry holds, the 
DOS diverges. Finally, we also consider the quasiparticle interaction. 
Consistently 
with the charge not being a conserved quantity, we find that the singlet 
particle-hole channel does not contribute, while the triplet particle-hole and 
the Cooper channels have a delocalizing effect, which also enhances the DOS. 
 
We consider the following Hamiltonian on a square lattice  
\begin{eqnarray}
{\cal H} &=& -\sum_{\langle ij\rangle} 
\sum_\sigma \left( t_{ij}{\rm e}^{i\phi_{ij}}
c^\dagger_{i\sigma}c^{\phantom{\dagger}}_{j\sigma} + H.c.\right)
\nonumber\\ 
&&+ \sum_{\langle ij\rangle} \left[\Delta_{ij} 
\left(c^\dagger_{i\uparrow}c^\dagger_{j\downarrow} 
+ c^\dagger_{j\uparrow}c^\dagger_{i\downarrow}\right) + H.c.\right],
\label{Hamiltonian}
\end{eqnarray}
where $\langle ij \rangle$ means that the sum is restricted to 
nearest neighbor sites, $c^\dagger_{i\sigma}$ creates an electron with 
spin $\sigma=\uparrow,\downarrow$ at site $i$, while 
$c^{\phantom{\dagger}}_{i\sigma}$ annihilates it. We take a 
gap function $\Delta_{ij}$ of $d$-wave symmetry. 
The hopping 
matrix elements are independent random 
variables and satisfy $t_{ij} = t_{ji}\in {\cal R}e$, 
and $\phi_{ij}=-\phi_{ji}$, with $\phi_{ij}$ zero or finite depending 
whether or not time reversal invariance holds.
The spectrum of the Hamiltonian (\ref{Hamiltonian}) 
posseses a nesting property.  
Namely, given a generic eigenfunction with 
energy $E$ and amplitude $\phi(i)_{E}$ at site $i=(n,m)$, the 
operator ${\cal O}_\pi \phi(i)_{E}\equiv (-1)^{n+m}\phi(i)_{E}$, which 
shifts by $(\pi,\pi)$ the momentum,    
generates the eigenfunction with 
energy $-E$. This implies an additional symmetry (chiral symmetry) at $E=0$.  
The localization properties are quite different whether 
chiral symmetry holds, which corresponds to the Fermi 
energy $E_F=0$ (half-filling), or broken, $E_F\not = 0$. 
In the latter case, the localization properties of   
(\ref{Hamiltonian}) are analogous to models in which   
on-site disorder is present or next-nearest neighbor hopping is 
included, which break chiral symmetry everywhere in the spectrum.

We analyse the disordered Hamiltonian (\ref{Hamiltonian}) by using the 
replica trick method within the path integral formalism\cite{EL&K}. 
We introduce the vector Grassmann variables  
$c_i$ and $\bar{c}_i$ with components 
$c_{i,\sigma,p,a}$ and 
$\bar{c}_{i,\sigma,p,a}$, where $i$ refers to a lattice site, $\sigma$ to the 
spin, $p=\pm$ is the index of positive ($+\omega$) 
and negative ($-\omega$) frequency components, 
and $a=1,\dots,n$ is the replica index, as well as the 
Nambu spinors 
\[
\Psi_i = \frac{1}{\sqrt{2}}
\left(
\begin{array}{c}
\bar{c}_i \\
i\sigma_y c_i\\
\end{array}
\right),
\]
and $\bar{\Psi}_i=\left[C\Psi_i\right]^t$, with the charge conjugation matrix 
$C=i\sigma_y \tau_1$. Here and in the following, 
the Pauli 
matrices $\sigma_b$ ($b=x,y,z$) act on the spin components, 
$s_b$ ($b=1,2,3$) on the frequency (retarded/advanced) components, 
and $\tau_b$ ($b=1,2,3$) on the Nambu components $\bar{c}$ and 
$c$. The action corresponding 
to (\ref{Hamiltonian}) is 
\begin{eqnarray}
S &=& \sum_{ij} \bar{\Psi}_i
\left( -t_{ij}{\rm e}^{-i\phi_{ij}\tau_3} 
+ i\Delta_{ij}\tau_2 s_1 -i\delta_{ij} \omega s_3\right)
\Psi_j 
\label{S}
\end{eqnarray}
As in the standard Abrikosov and Gorkov approach to 
superconductivity, the gap function couples fermions with   
opposite frequency $\omega$. 
If magnetic impurities are present, we must add to (\ref{S}) an 
additional spin-flip scattering term 
$\sum_i \bar{\Psi}_i \tau_3 \vec{\sigma}\cdot \vec{S}_i 
\Psi_i$, being $\vec{S}_i$ the impurity spin. The same term, with 
$\vec{S}_i = \vec{B}$, gives the Zeeman splitting in the 
presence of a constant magnetic 
field, which also breaks time-reversal invariance.     
From the action (\ref{S}) we can deduce the proper symmetries of the 
field theory. We consider two different global unitary symmetry 
transformations, 
one for sublattice $A$ and another for $B$, namely 
$\Psi_A \rightarrow T_A\Psi_A$ and $\Psi_B \rightarrow T_B\Psi_B$. For frequency $\omega=0$, 
the action is invariant if $CT_A^t C^t H_{AB} T_B = H_{AB}$. By writing 
$ T_A = \exp \frac{W_0+W_3}{2},\;\;
T_B = \exp \frac{W_0-W_3}{2}$,
with antihermitean $W$'s, we find that 
$
CW_0^t C^t = -W_0$ and   
$CW_3^t C^t = W_3$ ,
as well as 
$
\left[\tau_2 s_1, W_{0(3)}\right] =0$ and 
$\left[\tau_3, W_{0(3)}\right] =0$, 
the latter condition valid if time reversal invariance is broken. 
If magnetic impurities are present, then 
$\left[\tau_3\vec{\sigma}, W_{0}\right] =
\left\{\tau_3\vec{\sigma}, W_{3}\right\} =0$, while 
$\left[\tau_3\vec{\sigma}\cdot\vec{B}, W_{0}\right] =
\left\{\tau_3\vec{\sigma}\cdot\vec{B}, W_{3}\right\} =0$ in 
the presence of a constant magnetic field. 
Since the frequency $\omega$ acts as a symmetry breaking field, we have to further impose 
that $\left\{ W_0,s_3\right\} =0$ and 
$\left[W_3,s_3\right] =0$.
Finally, if chiral symmetry is broken by terms which couple same sublattices, 
we must set $W_3=0$. 
All the above conditions imply that 
the unitary matrices $T$ belong to a group G if $\omega=0$ which is 
lowered to a group H at $\omega\not = 0$.
In Table \ref{Table} we list the coset spaces G/H for the different classes
({\sl i}) time reversal invariance  holds with chiral symmetry \cite{Fukui2} 
or without \cite{Fisher,SQHE};
({\sl ii}) time reversal symmetry is broken with chiral symmetry \cite{Fukui2} 
or without \cite{Fisher};
({\sl iii}) a magnetic field is applied in the presence of chiral symmetry  
or without it \cite{Fisher,SQHE}; 
and finally ({\sl iv}) in the presence of magnetic impurities 
with chiral symmetry  
or in its absence\cite{Altland}.
Indeed, already at this 
stage we could anticipate the appropriate $\beta$-functions assuming that 
the diffusive modes are described by a non linear $\sigma$-model with the 
proper symmetry. However, for sake of clarity, we give here 
a sketchy derivation of the $\sigma$ model for the 
simple case of real hopping\cite{Luca}.
     
Within the replica trick method, we can average e$^{-S}$ over disorder. The 
effective action $S=S_0+S_{imp}$ is the sum of a regular part
$S_0$ given by Eq.(\ref{S}) with $t_{ij}=t$ (and $\phi_{ij}\equiv 0$)
plus the impurity contribution    
$S_{imp} = -2u^2 t^2 \sum_{\langle ij \rangle} 
\left(\bar{\Psi}_i\Psi_j\right)^2$.
Here $t$ is the average value of the nearest neighbor hopping, while $ut$ is the variance.   
We notice that $\bar{\Psi}_i\Psi_j = \bar{\Psi}_j\Psi_i$, so that, 
by introducing 
$X^{\alpha\beta}_i = \Psi^\alpha_i \bar{\Psi}^\beta_i$,
where $\alpha$ and $\beta$ is a multilabel for the different components, 
we can write
$
S_{imp} = 2u^2 t^2 \sum_{\langle ij \rangle} 
X^{\alpha\beta}_i X^{\beta\alpha}_j 
= 2u^2 t^2 \sum_{\langle ij \rangle} Tr\left(X_i X_j\right)
$.
In Fourier components 
\begin{equation}
S_{imp} = \frac{1}{V}\sum_{q\in BZ} W_q Tr\left(X_q X_{-q}\right),
\label{Simp}
\end{equation}
where $BZ$ means the Brillouin zone and 
$W_q = 2u^2 t^2\left(\cos q_x a + \cos q_y a \right)$,
$a$ being the lattice spacing. We can decouple (\ref{Simp}) by an 
Hubbard-Stratonovich transformation, introducing an auxiliary field. 
However, since $W_q=-W_{q+(\pi,\pi)}$ and 
$W_q>0$ if $q$ is restricted to the magnetic Brillouin zone ($MBZ$), we 
need to introduce two auxiliary fields defined within the $MBZ$, 
$Q_{0q}=Q_{0-q}^\dagger$ and $Q_{3q}=Q_{3-q}^\dagger$\cite{Fabrizio}, 
through which  
\begin{eqnarray}
S_{imp} &=& \frac{1}{V} \sum_{q\in MBZ} \frac{1}{4W_q} 
Tr\left[Q_{0q}Q_{0-q}+Q_{3q}Q_{3-q}\right] \nonumber\\
&-&
\frac{i}{V}\sum_{q\in MBZ} Tr\left[ Q_{0q}X^t_{-q} 
+ i Q_{3q}X^t_{-q-(\pi,\pi)}\right].
\label{Simpurity}
\end{eqnarray}
The above expression shows that $Q_0$ corresponds to smooth 
fluctuations of the auxiliary field, while $Q_3$ to staggered fluctuations. 
Namely, in the long-wavelength limit, the auxiliary field in real space is  
$Q_R = Q_{0R} + i(-1)^R Q_{3R}$, and it is not hermitean.     
The $\sigma$ model is derived by expanding the effective action for 
the auxiliary field $Q$, which is 
obtained after integration over the Grassmann variables, around the 
uniform saddle point in the presence of an infinitesimal 
symmetry breaking frequency.  Quite generally, 
the saddle point $Q_{sp}$ may have both a $\tau_0 s_3$ and a $\tau_2 s_1$ 
components, $\Sigma$ and $F$ respectively. However, for a $d$-wave order 
parameter, $F=0$, and $\Sigma$ satisfies
$1= \frac{u^2 t^2}{2}  
\frac{1}{V}\sum_k \frac{1}{E_k^2 
+ \Sigma^2}$,
where $E_k=\sqrt{\epsilon_k^2 + \Delta_k^2}$, 
$\epsilon_k=-2t(\cos k_x a + \cos k_y a)$ is the band energy, and 
$\Delta_k = \Delta\left[\cos(k_x a)-\cos(k_y a)\right]$ is the $d$-wave gap 
function. 
The above self-consistency equation leads to a finite density of state 
$N_0=\Sigma/(\pi u^2 t^2)$ at the chemical potential. 
We can write the $Q$-matrix through
$
Q_R = C T_R^t C^t \left(Q_{sp}+P_R\right)T_R\equiv 
Q(R) + C T_R^t C^t P_R T_R,
$
where $T_R$ belong to the coset G/H, and describe the transverse 
massless modes, 
while $P_R$ parametrize the longitudinal massive modes. The 
long wavelength action for the transverse modes is then obtained by integrating out 
the massive fluctuations and expanding in $\vec{\nabla} Q(R)$. 
As expected, we obtain a non linear $\sigma$-model type 
of action 
\begin{eqnarray}
S[Q] &=& \frac{\pi}{32\Sigma^2} \sigma 
\int dR\, Tr\left(\vec{\nabla} Q(R)^\dagger\cdot\vec{\nabla} Q(R)\right)
\nonumber\\
&&- 
\frac{\pi}{8\cdot 32 \Sigma^4} \Pi 
\int dR\, \left|Tr\left[Q(R)^\dagger \vec{\nabla} Q(R)\right]\right|^2,
\label{NLSM}
\end{eqnarray}
where 
$
\sigma = \frac{\Sigma^2}{\pi V}\sum_k \frac{ 
\vec{\nabla}\epsilon_k\cdot \vec{\nabla}\epsilon_k 
+ \vec{\nabla}\Delta_k\cdot \vec{\nabla}\Delta_k }
{\left(E_k^2+\Sigma^2\right)^2},
$
can be interpreted as the quasiparticle {\sl conductance} in the Drude 
approximation, which 
corresponds to the spin 
conductance of the system\cite{Fisher}. 
$\Pi$ is a parameter 
related to the staggered density of states fluctuations\cite{Fabrizio}. 
The last term only exists if chiral symmetry is present. 
The $\beta$-functions for the coupling constant $g=1/(2\pi^2\sigma)$ 
and for the DOS  
in the different universality classes are known\cite{Hikami}, and 
are shown in Table \ref{Table}. 
The RG equation for $c=1/(2\pi^2 \Pi)$ when chiral symmetry 
holds can be obtained by noticing that 
$\sigma + n\Pi$ is not renormalized\cite{Gade,Fabrizio}.  
Therefore $\beta(c)=-c^2\beta(g)/(ng^2)$. 
In Table \ref{Table} we define $\Gamma=g/(c+ng)$. The scaling behavior of the 
DOS are extracted from the one loop correction to $Q(R)$. 
 
According to Table \ref{Table}, in the zero replica limit we obtain that, 
if chiral symmetry is absent and for non magnetic impurities, the  
conductance vanishes, 
and the DOS, which is finite within the simplest Born 
approximation, is suppressed. As shown by Ref. \cite{FisherDOS}, 
in the localized phase the DOS vanishes as $|E|$ or 
$E^2$ depending whether time reversal symmetry holds or not.      
Quite surprisingly, magnetic impurities give a delocalization 
correction to the conductance, as well as a DOS enhancement. 
Finally, the symmetry class in the presence of a weak magnetic field 
is the same as in the standard Quantum Hall effect\cite{Fisher,SQHE}.  

On the contrary, if chiral symmetry is present, the conductance stays 
finite, or even increases in the presence of a 
magnetic field. Without magnetic field and in the absence of spin flip scattering, 
the DOS diverges approximately like 
$\rho(E)\sim \exp\left[A\sqrt{-\ln E}\right]/E$, with $A$ a model dependent 
constant\cite{Gade,Fabrizio}. By studying an on-site disordered model 
in the unitary scattering limit, a 
$\rho(E)\sim 1/(E\ln^2 E)$ was found in 
Ref. \cite{Pepina}. Since  in the unitary limit, the on-site 
disorder effectively reduces to a random hopping, the two results should 
coincide. Indeed the leading divergence $1/E$ is the same, while the 
subleading behavior is not, reflecting the different way of summing up 
the singularities induced by disorder. 

The quasiparticle charge modes, as well as the spin modes   
when magnetic impurities or a magnetic field are present, 
are not described by the non linear $\sigma$-model (\ref{NLSM}), 
which only 
represents the truly massless diffusion modes. Nevertheless, charge and 
spin conductivities, 
$\sigma_c$ and $\sigma_s$, respectively,  can be still evaluated through 
the stiffness of the 
corresponding modes, although they acquire a mass term. 
We have checked that 
this procedure gives the same results as the explicit evaluation 
of the one loop corrections to the $Q$-field operators which describe the charge and spin 
conductivities by introducing 
sources which directly couple 
to the currents \cite{Castellani}. 
In this way we find that the one 
loop corrections $\delta \sigma_c/\sigma_c$ and 
$\delta \sigma_s/\sigma_s$ coincide with 
$\delta \sigma/\sigma$ in the absence of sublattice symmetry.
On the contrary, when sublattice symmetry holds, quasiparticle charge 
conductivity is reduced at one loop order and in the zero replica limit 
according to  
$
\delta \sigma_c = -\sigma_c 2 g \ln s
$
(where $s$ is the momentum rescaling factor),
unless spin flip scattering is  present, in which case $\delta\sigma_c=0$. 
Moreover the spin conductance renormalizes as $\sigma$ 
in the absence of magnetic field and magnetic impurities, 
while $\delta\sigma_s=0$ otherwise. 
Notice that  while the bare $\sigma_s$ 
coincides with the bare $\sigma$, the bare $\sigma_c$ 
is given by a similar expression as the bare $\sigma $ but without  
the current vertex $\vec{\nabla}\Delta_k$\cite{Durst&Lee}. 

We can introduce the residual quasiparticle interaction, which we assume 
is repulsive,  
in the spirit of the Landau Fermi liquid theory as originally done 
by Finkelstein in the standard Anderson localization\cite{Finkelstein}.  
Interaction generically spoils sublattice symmetry already at the level of 
the Fock's diagrams, unless for particular cases like on-site or nearest 
neighbor site interactions. Here we consider a generic 
case without sublattice symmetry, where   
just the three quasiparticle scattering amplitudes considered in Ref. 
\cite{Finkelstein} are important: 
the particle-hole (p-h) scattering amplitudes 
in the singlet, $U_s$, and triplet, $U_t$, channels, 
and the particle-particle (p-p) channel scattering amplitude, $U_c$. 
Being the charge not a conserved quantity, 
the singlet p-h channel has no component on the transverse massless modes, 
contrary to the p-h triplet channel. The 
p-p channel couples Cooper pairs of $s$-wave symmetry. Since the 
real part of an $s$-wave order parameter is in competition with the 
$d$-wave symmetry, only 
the p-p $\tau_1$ channel contributes, which corresponds    
to fluctuations of an $is$ pairing  
order parameter. 
Moreover, their contribution to $\sigma$ and 
to the DOS is exactly the same as in the absence of superconductivity, apart 
from a factor $1/2$ in $U_c$ being absent the p-p $\tau_2$-channel. 
Following Ref. \cite{Finkelstein}, we find 
the first order interaction corrections  
\begin{equation}
\frac{\delta\sigma}{\sigma_0} =
\frac{\delta N}{N_0} = 2gN_0\left(\frac{3}{2}U_t+\frac{1}{2}U_c\right)\ln s.
\label{interaction}
\end{equation}
If time reversal invariance is broken, the $U_c$ contribution 
drops out. A Zeeman term removes 2/3 of the triplet contribution, 
while in the presence of magnetic impurities the whole correction 
vanishes. Therefore a repulsive interaction has a delocalizing effect which 
competes with quantum interference corrections. Moreover, both 
the p-h triplet channel and the p-p $is$-channel\cite{Khveshchenko}  
enhance the DOS.  
However, being the 
interaction correction proportional to the small DOS induced by disorder, 
it is likely negligible, although it might become dominant at very low 
temperatures if the enhancement of the p-h triplet  
scattering amplitude takes place\cite{Finkelstein}, while the p-p one is  
suppressed at higher energy by the Cooper 
phenomenon\cite{Finkelstein,Khveshchenko}. For large values of the triplet
amplitude Eq.(\ref{interaction}) is no more appropriate and a RPA resummation
of the dynamical amplitude should be carried out leading to the same expressions
for $\delta\sigma/\sigma_0$ and $\delta N/N_0$ as in the absence of 
superconductivity \cite{Finkelstein}.

Finally, let us compare the above theoretical predictions with 
the experimental behavior of the superconducting cuprates. 
In the realistic situation in which time reversal symmetry is unbroken 
and chiral symmetry is absent, the 
quantum interference corrections should appear when 
$\ln \tau_\phi/\tau \simeq (v_F^2+v_\Delta^2)/v_Fv_\Delta$, where 
$\tau=1/2\Sigma$ is the elastic relaxation time induced by disorder, 
$\tau_\phi$ the inelastic one, $v_F$ and $v_\Delta$ the velocities 
parallel and perpendicular to the nodal directions. 
Due to the still controversial situation for the inelastic relaxation time in BSCO, 
let us estimate the crossover temperature $T_*$ 
below which localization effects should appear in YBCO. On the basis of the 
microwave conductivity data of Ref. \cite{Hosseini} for optimally 
doped YBCO, one finds $T_*\sim 1$ K if $\tau_\phi \propto T^{-4.2}$ 
is assumed\cite{Hosseini}, or $T_* \sim 0.3$ K for 
$\tau_\phi \propto T^{-3}$\cite{Scalapino}. Moreover, $T_*$ might be further reduced 
either by interaction and nesting effects, as shown previously, or 
by a finite correlation range of the impurity potential\cite{Shura,Altland}.  
Experimentally, optimally doped 
YBCO and BSCO thermal conductivities \cite{Taillefer,Chiao} 
do not show any localization 
suppression down to 0.1 K, which is however  not so small as compared to our 
estimates for $T_*$ to exclude that localization finally takes place.     
Indeed in underdoped YBCO there is evidence of a vanishing 
residual quasiparticle conductivity, as extracted by heat transport
\cite{Hussey}.

In summary, we have derived the general form of the non linear 
$\sigma$-model, which describes the quasiparticle diffusive modes in a 
$d$-wave supercondutor, in the presence of a random hopping. 
Away from half-filling, the random hopping plays a similar role as 
a random on-site potential.  However, at half-filling, the model 
posseses an additional chiral symmetry and, as a result, it describes 
a critical theory with a finite spin conductance. 
We believe that the results are of more general validity than the 
peculiar choice of the disorder would suggest.  
Indeed the model is also applicable in the case of a strong 
disordered impurity potential close to the unitary scattering limit, which 
effectively reduces to a random hopping. Incidentally, 
this is believed to be just the case for the High T$_c$ superconductors.  
We have also calculated the localization corrections to the 
quasiparticle charge conductivity, namely of the optical conductivity 
at small frequency, and find that it should vanish at sufficiently 
low temperature in all cases but in the 
presence of magnetic impurities. 

Finally, we have included in our analysis a residual quasiparticle 
interaction, which we have shown has a delocalizing effect mainly due 
to the quasiparticle spin diffusive modes. The results suggest that 
these systems, although being singlet superconductors, might support  
consistent spin fluctuations.  

{\sl Note added } While writing this paper, we became aware that 
Jeng, Ludwig, Senthil and Chamon \cite{Jeng} 
have studied the interaction effects 
in the same model within the Keldisch technique, obtaining results similar to 
ours.

We would like to acknowledge helpulf discussions with A.A. Nersesyan. 
We also thank A.W.W. Ludwig for having told us the results obtained in Ref. 
\cite{Jeng} prior of its publication.

\begin{table}
\caption{Coset spaces and $\beta$ functions for the 
coupling $g$ and for the DOS $N$ in the different universality classes.
 $\hat{T}$ is the time reversal invariance.}
  
\begin{tabular}{|c||c|c|c|} 

~&  Coset space &  $\beta_g$       &   $\beta_N$  \\ \hline\hline 
 
Yes Chiral, Yes $\hat{T}$ & U$(4n)\times$U$(4n)$/U$(4n)$  
& $8ng^2$ & $(\Gamma/4-8n)g$  \\ \hline

Yes Chiral, No $\hat{T}$ &  U$(4n)$/O$(4n)$  
&         $4ng^2$       &   $(-1+\Gamma/4-4n)g$  \\ \hline

Yes Chiral, magnetic field &  O$(4n)$/O$(2n)\times$O$(2n)$ &
$(2n-1)g^2$  & $-2ng$ \\ \hline

Yes Chiral, spin flip & U$(2n)$/U$(n)\times$U$(n)$ &
$ng^2$ & $-ng$ \\ \hline

No Chiral, Yes $\hat{T}$ & Sp$(2n)\times$Sp$(2n)$/Sp$(2n)$
& $2(2n+1)g^2$ & $(-1-4n)g$   \\ \hline

No Chiral, No $\hat{T}$  & Sp$(2n)$/U$(2n)$
&    $(2n+1)g^2$         &   $(-1-2n)g$ \\ \hline

No Chiral, magnetic field & U$(2n)$/U$(n)\times$U$(n)$           
&      $ng^2$       & $-ng$  \\ \hline

No Chiral, spin flip & O$(2n)$/U$(n)$                
&   $(n-1)g^2$    & $(1-n)g$  \\ 
\end{tabular}
\label{Table}
\end{table}


\begin{references}
\bibitem{Luca} A detailed derivation will be presented elsewhere.
\end{references}
\end{document}